\documentclass[12pt]{article}
\usepackage{graphicx}
\usepackage{xspace}
\usepackage{gensymb}
\usepackage[gen]{eurosym}
\usepackage{amsmath} 
\usepackage{amssymb} 
\usepackage[english]{babel}
\usepackage[left]{lineno}

\newcommand{\nubar}{$\overline{\nu}\ $}
\newcommand{\nue}{\ensuremath{\nu_{e}}\xspace}
\newcommand{\nubare}{\ensuremath{\overline{\nu}_{e}}\xspace}

\newcommand{\numu}{\ensuremath{\nu_{\mu}}\xspace}
\newcommand{\nust}{\ensuremath{\nu_s}\xspace}
\newcommand{\nubarmu}{\ensuremath{\overline{\nu}_{\mu}}\xspace}

\newcommand{\nutau}{\ensuremath{\nu_{\tau}}\xspace}
\newcommand{\numunue}{\ensuremath{\nu_\mu \rightarrow \nu_e}\xspace}
\newcommand{\numunumu}{\ensuremath{\nu_\mu \rightarrow \nu_\mu}\xspace}
\newcommand{\numunust}{\ensuremath{\nu_\mu \rightarrow \nu_s}\xspace}
\newcommand{\nuenue}{\ensuremath{\nue \rightarrow \nue}\xspace}

\newcommand{\numunutau}{\ensuremath{\numu \rightarrow \nutau}\xspace}

\newcommand\pubnumber{SNSN-323-63}
\newcommand\pubdate{\today}

\def\padova{INFN Padova\\
Via Marzolo, 8, I-35131 Padova, ITALY}

\def\Title#1{\begin{center} {\Large #1 } \end{center}}
\def\Author#1{\begin{center}{ \sc #1} \end{center}}
\def\Address#1{\begin{center}{ \it #1} \end{center}}

\newcommand\pubblock{\rightline{\begin{tabular}{l} \pubnumber\\
         \pubdate  \end{tabular}}}
\newenvironment{Abstract}{\begin{quotation}  }{\end{quotation}}
\newenvironment{Presented}{\begin{quotation} \begin{center} 
             PRESENTED AT\end{center}\bigskip 
      \begin{center}\begin{large}}{\end{large}\end{center} \end{quotation}}
\def\Acknowledgements{\bigskip  \bigskip \begin{center} \begin{large}
             \bf ACKNOWLEDGEMENTS \end{large}\end{center}}

\def\beq{\begin{equation}}
\def\eeq#1{\label{#1}\end{equation}}
\def\eeqn{\end{equation}}

\def\beqa{\begin{eqnarray}}
\def\eeqa#1{\label{#1}\end{eqnarray}}
\def\eeqan{\end{eqnarray}}

\let\bar=\overbar

\def\Dslash{\not{\hbox{\kern-4pt $D$}}}
\def\dslash{\not{\hbox{\kern-2pt $\del$}}}

\def\msb{{\bar{\ssstyle M \kern -1pt S}}}

\begin{document}
\begin{titlepage}
\pubblock

\vfill
\Title{Search for Sterile Neutrinos at Long and Short Baselines}
\vfill
\Author{ Luca Stanco
}
\Address{\padova}
\vfill
\begin{Abstract}
Neutrino physics is currently suffering from lack of knowledge from at least four major ingredients.
One of them is the presence or not of new sterile neutrino states at the mass scale of around 1 eV. Settling this point
should be the highest priority for the neutrino community. We will discuss the state--of--the art of
experimental searches for sterile neutrinos with accelerators, both at long and short baselines.
\end{Abstract}
\vfill
\begin{Presented}
NuPhys2015, Prospects in Neutrino Physics\\
Barbican Centre, London, UK,  December 16--18, 2015
\end{Presented}
\vfill
\end{titlepage}
\def\thefootnote{\fnsymbol{footnote}}
\setcounter{footnote}{0}

\section{Introduction}

The current scenario of the Standard Model (SM) of particle physics, being arguably stalled by the discovery of the Higgs boson,
is {\em desperately} looking for new experimental inputs to provide a more comfortable and {\em conformable} theory.
In parallel, experiments on neutrinos so far have been  an outstanding source of novelty and unprecedented 
results. In the last two decades several results were obtained 
by studying atmospheric, solar or reactor neutrinos, or more recently with neutrinos 
from accelerator--based beams. Almost all these results have contributed to strengthen 
the flavour--SM. 
Nevertheless, relevant parts like the values of the leptonic CP phase, $\delta_{CP}$ and the neutrino masses are still missing, a critical ingredient
being the still undetermined neutrino mass ordering.

Furthermore, tensions still exist among some experimental results. As shown below, the most relevant concerns
the existence or not of a non--standard neutrino state at the mass scale of 1 eV. These states, ``sterile neutrinos''
in the original definition of Bruno Pontecorvo in 1968~\cite{pontecorvo}, reached the level of suggestive possibility since in the last years
three different kinds of experiments hinted at their existence. 
The excess of \nue (\nubare) observed originally by the LSND~\cite{lsnd} collaboration and the (inconclusive) result by 
MiniBooNE~\cite{miniB} collaboration, as well as the so-called 
reactor~\cite{reactor} and Gallium~\cite{gallium1, gallium2} neutrino anomalies can be coherently interpreted as due to the existence 
of at least a fourth sterile neutrino with a mass at the eV scale.

As a matter of fact both $\delta_{CP}$ and the mass ordering, together with the technical but relevant ingredient of the still undetermined
amount of the deviation, with sign, of the {\em atmospheric} mixing angle, $\theta_{23}$, from $\pi/4$, are
tightly inter--connected to get a comprehensive description of the neutrino oscillation paradigm. As demonstrated
very recently~\cite{palazzo-2015} these three parameters may depend on the further occurrence of neutrino sterile states at the 1 eV mass scale.
Therefore, a clarification of the sterile issue is mandatory. A more detailed discussion of the past and current scenario can be found 
in~\cite{stanco-next}.

The most critical point in the search for sterile states stays in the lack of any appearance/disappearance affecting \numu or \nutau neutrinos.
The set of results that addresses the neutrino anomalies read off by LSND {\em et al.} are all related to appearance/disappearance effects
for \nue and \nubare. There are actually phenomenological strong tensions when $3 + 1$ or $3 + n$ models are used to 
interpret the \nue (\nubare) appearance/disappearance and the corresponding, required, appearance/disappearance of the other flavours, 
\numu and \nutau~\cite{tension}.

In this report we will focus on the current experimental activity about the search for sterile neutrinos at 1 eV, based on accelerator
\numu beams, either in a long--baseline or a short--baseline configuration.

\section{Sterile neutrinos}

The presence of an additional sterile--state can be expressed in the extended PMNS~\cite{pontecorvo,pmns} mixing matrix 
($U_{\alpha i}$ with $\alpha = e, \mu, \tau, s$, and $i = 1,\ldots,4$). 
In this model, called ``3+1'', the neutrino mass eigenstates $\nu_1,\ldots,\nu_4$ are labeled such that the first three states are mostly made 
of active flavour states and contribute to the ``standard'' three flavour oscillations with the squared mass differences 
$\Delta\, m_{21}^2 \sim 7.5\times 10^{-5}~{\rm eV^2}$ and $|\Delta\, m_{31}^2| \sim 2.4\times 10^{-3}~{\rm eV^2}$, 
where $\Delta\, m_{ij}^2 = m^2_i - m^2_j$. The fourth mass eigenstate, which is mostly sterile, is assumed to be much heavier than the others, 
$0.1~{\rm eV^2}\lesssim \Delta\, m_{41}^2 \lesssim 10~{\rm eV^2}.$ The opposite case in hierarchy, i.e. negative values of $\Delta\, m_{41}^2$, produces a similar phenomenology from the 
oscillation point of view but is disfavored by cosmological results on the sum of neutrino masses~\cite{cosmo-data}\footnote{Actually
cosmology tends more and more even to disprove the NH case and therefore the existence of a relativistic specie at 1 eV mass.}. 

The phenomenology at short--baseline (SBL) is simplified since $L/E\sim 1$ km/GeV. Thus
the oscillation effects due to $\Delta m^2_{21}$ and $\Delta m^2_{31}$ can be neglected. 
Therefore the oscillation probability depends only on $\Delta m^2_{41}$ and $U_{\alpha 4}$ with $\alpha = e,\mu,\tau$.
In particular the survival probability of muon neutrinos 
is given an the effective two--flavour oscillation formula:
\begin{equation}\label{eq:1}
P(\nu_{\mu}\to\nu_{\mu})_{SBL}^{3+1} = 1 - \left[ 4 \vert U_{\mu 4} \vert^2 (1 - \vert U_{\mu 4} \vert^2)\right]  \sin^2 \frac{\Delta m^2_{41} L}{4E},
\end{equation}
where 
$4 \vert U_{\mu 4} \vert^2 (1 - \vert U_{\mu 4} \vert^2)$ is the {\em amplitude} and, since the baseline $L$ is fixed by the experiment location, 
the oscillation {\em phase} is driven by the neutrino energy E.

In contrast, appearance channels (i.e. $\nu_\mu \to \nu_e$) are driven by
terms that mix up the couplings between the initial and final flavour--states and
the sterile state, yielding a more complex picture:
\begin{equation}\label{eq:2}
P(\nu_{\mu}\to\nu_e)_{SBL}^{3+1} = 4 \vert U_{\mu 4}\vert^2 \vert U_{e 4} \vert^2  \sin^2 \frac{\Delta m^2_{41} L}{4E}
\end{equation}
Similar formulas hold also assuming more sterile neutrinos ($3 + n$ models).

Since $\vert U_{\alpha 4}\vert$ is expected to be small, the appearance channel is suppressed by two more
powers in $\vert U_{\alpha 4}\vert$ with respect to the disappearance one. Furthermore, since $\nu_e$ or $\nu_\mu$ appearance
requires $\vert U_{e 4}\vert > 0$ and $\vert U_{\mu 4}\vert > 0$, it should be naturally accompanied by
non--zero $\nu_e$ and $\nu_\mu$ disappearances. In this sense the disappearance
searches are essential for providing severe constraints  on the theoretical models 
(a more extensive discussion on this issue can be found e.g. in Section~2 of~\cite{winter}).

It should be also mentioned that a good control of the $\nue$ contamination is
important when using the $\nu_\mu \to \nu_e$ for sterile neutrino searches at SBL.
In fact the observed number of  $\nu_e$ neutrinos would depend on both the $\nu_\mu\rightarrow\nu_e$ appearance
and the  $\nu_e\rightarrow\nu_s$ disappearance, from the \numu and \nue components of the beam, respectively. 
On the other hand, the amount of $\nu_\mu$ neutrinos would be affected  by the
$\nu_\mu\rightarrow\nu_s$  and $\nu_e\rightarrow\nu_\mu$ transitions. However the latter term (\numu appearance)  
 would be much smaller than in the \nue case since the $\nu_e$ contamination in $\nu_\mu$ beams is
usually at the percent level. 
We would thus conclude that oscillation probabilities in the $\nu_{\mu}$ disappearance channel, in either a near or a far detector,
are not affected by any interplay of different flavours. Since both near and far detectors measure the same individual
disappearance transition, the probability amplitude should be the same at both sites.

The situation is quite more complicated in the long--baseline (LBL) configuration.
When $L/E \gg 1$ km/GeV, that is the case for the Long-Baseline experiments, the two--flavour oscillation
is not a good approximation. In the case of the CNGS beam, when studying the
\nutau oscillation rate the only valid approximations correspond to neglect
 the solar--driven term, i.e. $\Delta\, m_{21}^2 \sim 0$,  and to discard  the \nue component of
the beam. However when the \numunue channel is studied the intrinsic \nue beam--component becomes 
a non--negligible factor~\cite{palazzo}.

Considering the (\numu, \nutau, $\nu_s$) triplet, together with the above two approximations, the oscillation probability \numunutau can be written as:
\begin{eqnarray*}{P_{\nu_\mu\to\nu_{\tau}} = {4 |U_{\mu 3}|^2|U_{\tau 3}|^2\sin^2\frac{\Delta_{31}}{2}} + {4 |U_{\mu 4}|^2|U_{\tau 4}|^2\sin^2\frac{\Delta_{41}}{2}}}\\
{ +\, { 2\Re[U^*_{\mu 4}U_{\tau 4}U_{\mu 3}U_{\tau 3}^*]\sin\Delta_{31}\sin\Delta_{41}}}\\
{ -\, { 4\Im[U^*_{\mu 4}U_{\tau 4}U_{\mu 3}U_{\tau 3}^*]\sin^2\frac{\Delta_{31}}{2}\sin\Delta_{41}}}\\
{ +\, { 8\Re[U^*_{\mu 4}U_{\tau 4}U_{\mu 3}U_{\tau 3}^*]\sin^2\frac{\Delta_{31}}{2}\sin^2\frac{\Delta_{41}}{2}}}\\
{+\, {4 \Im[U^*_{\mu 4}U_{\tau 4}U_{\mu 3}U_{\tau 3}^*]\sin\Delta_{31}\sin^2\frac{\Delta_{41}}{2}}},
\end{eqnarray*}
using the definition $\Delta_{ij}=1.27\; \Delta\, m^2_{ij}\; L/E$ (i,j=1,2,3,4), with $\Delta_{31}$ and $\Delta_{41}$
expressed in eV$^2$, $L$ in km and $E$ in GeV. The first term corresponds to the standard oscillation, the
second one to the pure exotic oscillation, while the last 4 terms correspond to the interference between the standard
and sterile neutrinos. By defining $C=2|U_{\mu3}||U_{\tau3}|$, 
$\phi_{\mu\tau}=Arg(U^{\star}_{\mu3}U^{\star}_{\tau3}U^{\star}_{\mu4}U^{\star}_{\tau4})$ and 
$\sin\, 2\theta_{\mu\tau}=2|U_{\mu4}||U_{\tau4}|$, and noting explicitly the dependence of the probability on energy $E$, 
the expression can be re--written as:
\begin{eqnarray*}
{P(E) = { C^2 \sin^2\frac{\Delta_{31}}{2}} + {\sin^2\, 2\theta_{\mu\tau}\sin^2\frac{\Delta_{41}}{2}}}\\
{ +\, { \frac{1}{2} C \sin\, 2\theta_{\mu\tau}\cos\phi_{\mu\tau}\sin\Delta_{31}\sin\Delta_{41}}}\\
{ -\, { C\sin2\theta_{\mu\tau}\sin\phi_{\mu\tau}\sin^2\frac{\Delta_{31}}{2}\sin\Delta_{41}}}\\
{ +\, { 2\, C\sin2\theta_{\mu\tau}\cos\phi_{\mu\tau}\sin^2\frac{\Delta_{31}}{2}\sin^2\frac{\Delta_{41}}{2}}}\\
{+\, {C\sin 2\theta_{\mu\tau}\sin\phi_{\mu\tau}\sin\Delta_{31}\sin^2\frac{\Delta_{41}}{2}}},
\end{eqnarray*}
where interesting dependences rise up, namely the sign of $\Delta\, m^2_{13}$ (3$^{rd}$ and 6$^{th}$ terms) and 
CP-violating terms (4$^{th}$ and 6$^{th}$ terms). Finally, since at LBL $L/E \gg 1$ one can average over
the energy obtaining $<\sin\Delta_{41}>\approx 0$
and $<\sin^2\frac{\Delta_{41}}{2}> \approx \frac{1}{2}$. The following expression is pulled out:
\begin{eqnarray*}
{P(E) \simeq { C^2 \sin^2\frac{\Delta_{31}}{2}} + {\frac{1}{2}\sin^22\theta_{\mu\tau}}}\\
{ +\, { C\sin2\theta_{\mu\tau}\cos\phi_{\mu\tau}\sin^2\frac{\Delta_{31}}{2}}}\\
{+\, {\frac{1}{2}C\sin\, 2\theta_{\mu\tau}\sin\phi_{\mu\tau}\sin\Delta_{31}}}.
\end{eqnarray*}
This formula indicates that we are sensitive to the effective sterile mixing angle, $\theta_{\mu\tau}$, the mass hierarchy (MH, Normal NH or Inverted IH) and to the new CP--violating phase. 

\section{OPERA preliminary results on sterile neutrinos from \nutau and \nue}\label{sect-3}

What occurs in the long--baseline scenario is an evident interplay of interference effects,
such that there are zones of phase--space ($\sin^2\, 2\theta_{\mu\tau}$, $\Delta\, m^2_{41}$)
where more events are expected and zones where less events are
expected. On top of that the mass hierarchy has to be disentangled.
Therefore the method carried out by the OPERA collaboration is independently applied to 
the NH and IH cases. Maximization of the likelihood is performed over $\phi_{\mu\tau}$, $C$ and $\theta_{\mu\tau}$, i.e
the CP-violation phase and the two effective mixing angles of the 3$^{rd}$ and $4^{th}$ 
mass--states 
with \numu and \nutau, respectively. 

\begin{figure}\begin{center}
     \includegraphics[width=.6\textwidth]{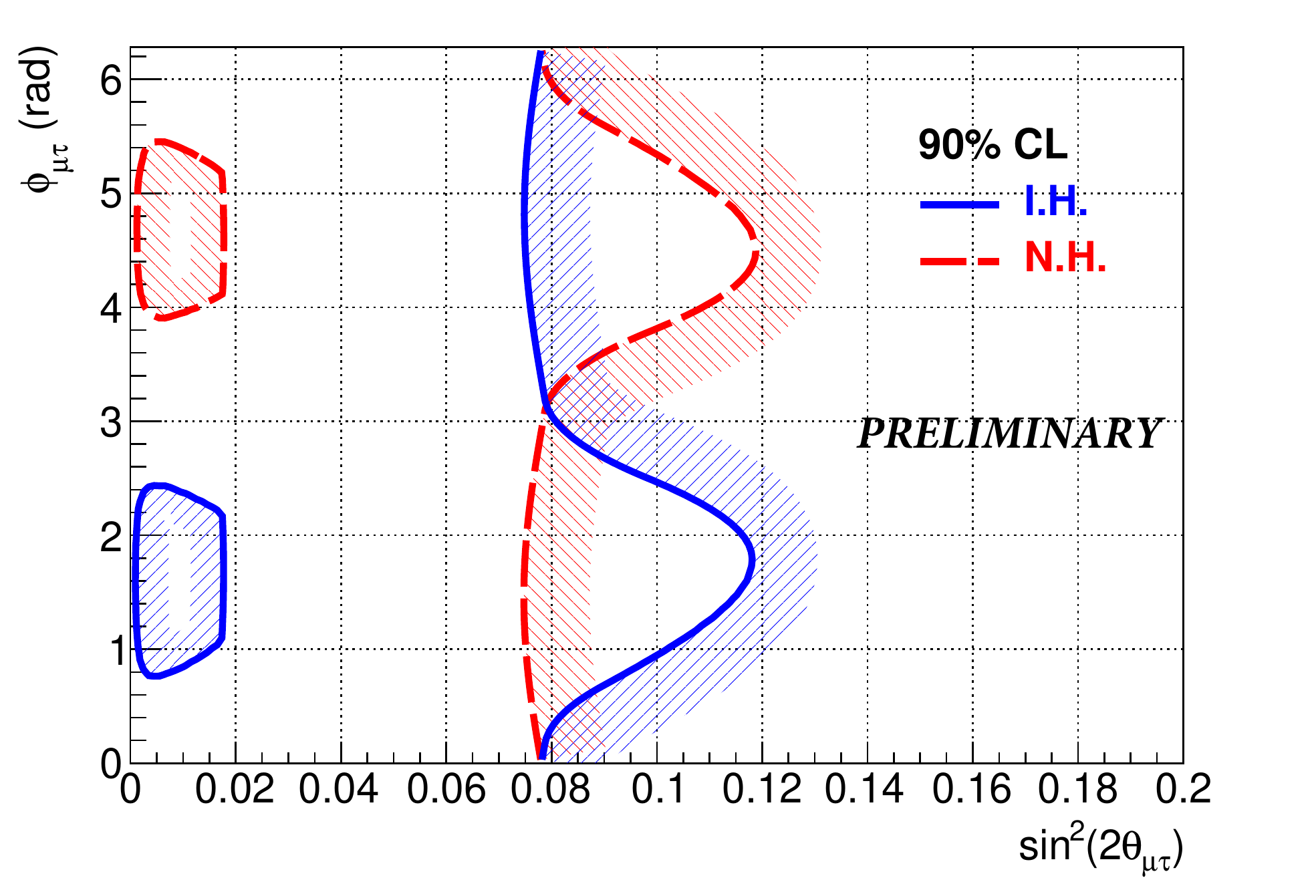}
     \caption{90\% C.L. exclusion limits in the $\phi_{\mu\tau}$ vs $\sin^2\, 2\theta_{\mu\tau}$ parameter space for normal 
(NH, dashed red) and inverted (IH, solid blue) hierarchies, assuming $\Delta\, m_{41}>1$ eV$^2$. Bands are drawn to 
indicate the excluded regions.}
     \label{fig2}
\end{center}\end{figure}

Results on sterile limits based on four \nutau candidates have been published by OPERA~\cite{opera-4th}.
An updated analysis based on the just discovered 5$^{th}$ candidate~\cite{opera-5th} was released last
year~\cite{stanco-opera}. 
Since OPERA sensitivity is limited to the region ($\sin^2\, 2\theta_{\mu\tau}\gtrsim 0.1$, 
$\Delta\, m_{41}\gtrsim 0.01$ eV$^2$) its analysis was two--fold. In the first case the
$\Delta\, m_{41}~>~1$~eV$^2$ region was considered, where almost no correlation with the effective mixing angle is 
present and the exclusion limit on the plane of the phase vs the mixing angle  can be extracted (figure~\ref{fig2}).
When marginalization over the phase is made, the limit $\sin^2\, 2\theta_{\mu\tau}< 0.11$ at 90\% C.L. is obtained
(preliminary).

\begin{figure}\begin{center}\vskip-0.7cm
     \includegraphics[width=.7\textwidth]{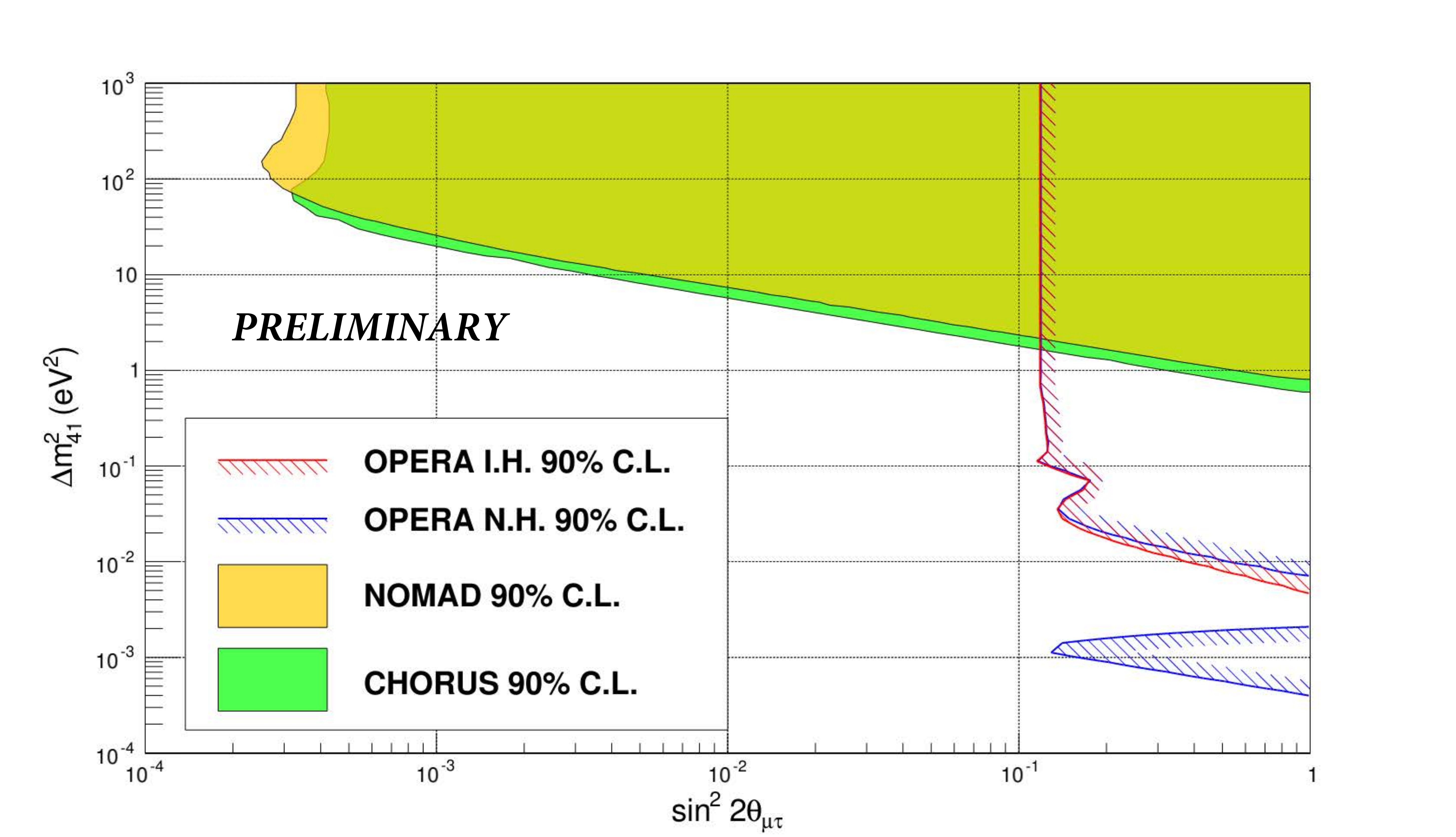}
     \caption{OPERA preliminary 90\% C.L. exclusion limits in the $\Delta\, m^2_{41}$ vs $\sin^2\, 2\theta_{\mu\tau}$ parameter space for
the normal (NH, red) and inverted (IH, blue) hierarchy of the three standard neutrino masses. The exclusion 
plots by NOMAD~\cite{nomad} and CHORUS~\cite{chorus} are also shown. Bands are drawn to indicate the excluded 
regions. It is intriguing that for NI
life seems harder than for IH.}
     \label{fig3}
\end{center}\end{figure}

To extend the search for a possible fourth sterile neutrino down to small $\Delta\, m^2_{41}$ values, a second
kind of analysis has been performed by OPERA using the GLoBES software,
which takes into account the non-zero $\Delta\, m^2_{12}$ value and also matter effects, the Earth density being approximated by a constant value estimated with the PREM shell--model.
This time the $\Delta\, m^2_{31}$ parameter has been profiled out (see~\cite{opera-4th} for more details and references).
In figure~\ref{fig3} the preliminary 90\% CL exclusion plot is reported in the $\Delta\, m^2_{41}$ vs $\sin^2\, 2\theta_{\mu\tau}$ 
parameter space. The most stringent limits of direct searches for \numunutau oscillations at short-baselines obtained 
by the NOMAD~\cite{nomad} and CHORUS~\cite{chorus} experiments are also shown.

Another very interesting analysis is in progress in OPERA on the \numunue search. The 2013 analysis~\cite{opera-old}
 will be updated
using the full data set and a much less approximate analysis as previously done. The preliminary selection is shown in 
table~\ref{tab1}. The exclusion region will be set in the plane $\Delta\, m^2_{41}$ vs $\sin^2\, 2\theta_{\mu e}$.

\begin{table}[h]
\begin{center}\vskip0cm
     \begin{tabular}{l|c|c}
      & all energy range & $E<20$ GeV \\ \hline
     \nue candidates (30\% data) & 19 & 4\\ 
     background (30\% data) & 19.2$\pm$ 2.8 & 4.6\\
     \nue candidates (all data), preliminary & 52 & 9\\ \hline
     \end{tabular}
\end{center}
     \caption{The published selected \nue candidates (and expected background) by OPERA, corresponding to the 30\% of data sample
      and the preliminary selection on the full statistics.}
     \label{tab1}
     \end{table}

\section{MINOS and SuperK analyses}\label{sect-4}

The MINOS and SuperK collaborations have also studied in detail the \numunumu and \nuenue oscillations to 
exclude extra contributions from \numunust oscillations. Recent results have been given by MINOS
that makes use of the NuMI beam at FNAL~\cite{minos}, and SuperK by using the atmospheric neutrino flux~\cite{superk}.
MINOS is also analyzing the \nubar running--mode data--sample and their updated analysis
on \numunue will be soon released. 

The SuperK analysis is two--fold, considering either  $|U_{e4}|=0$ with matter effects or the full PMNS
and discarding the matter effect. In the latter case a strong limit is obtained, $|U_{\mu4}|<0.04$ at 90\% C.L., for
$\Delta\, m^2_{41} > 0.1$ eV$^2$, for a total exposure of 282 kton--year.

\section{Short--baseline experiments}
 In the short--baseline configuration life is {\em a priori} much easier. The two--flavour approximation is valid
 (see equations~\ref{eq:1} and~\ref{eq:2}). However several concerns rise up. The right choices of detectors
 for more than one site, with an effective inter--calibration and overlap over a large interval of energies, to overcome
  either the systematical errors or the interplayed oscillations between all the components of the beam, is mandatory
  to achieve a full disentlanging.
 
  The new SBL project at Fermilab~\cite{fnal-sbl} owns an excellent program for the measurement of the still
   not well know cross--sections, with an excellent technology. However the use of only the technique of Liquid Argon,
    mainly due to better develop the detectors in view of the long--baseline experiment, DUNE, may not be sufficiently
    robust e.g. for the measurement of the \numu disappearance.
    The unique features of an experiment that would single out the behaviors of \numu and \nubarmu~\cite{nessie} will unfortunately 
     not be available as the proposal was not approved by FNAL.
     The NESSiE experiment, with its exploitation of the charge measurement on event--by--event basis, would be able to 
     provide a unique gain in sensitivity of more than one order of magnitude in the mixing angle, for both neutrino and
     anti--neutrino channels, also challenging the interpretation of the anomalies at 1 eV scale as due to an oscillation
      with new neutrino sterile states.

\section{Conclusions and perspectives}\label{sect-5}

The long--standing issue on the existence of sterile neutrino states at the eV mass scale can receive new relevant inputs 
from the accelerator Long--Baseline experiments, like OPERA, MINOS and SuperK. From one side LBL, owing
to the large $L/E$ values, can only look at the averaged oscillation pattern (lacking any oscillatory behavior of data).
From the other side, the not--negligible interference between flavours introduces dependencies on the mass hierarchy and the CP--violation phase. 

New results were recently
published by the three collaborations, either on \numunumu disappearance or on the \numunutau appearance. 
All the results put stringent exclusion limits on the effective mixing angles between \numu/\nutau and \nust, so increasing the
tension with the positive results on \nue appearance/disappearance. With regards to \nue, OPERA and MINOS$+$
will soon release reliable results with their large data--set, by properly taking into account the extended $3+1$ scenario.

In case of existence of a sterile neutrino at the eV mass--scale this situation points towards a rather low effective mixing angle,
of the order of 1\%, between sterile and the standard neutrino flavours. 
Therefore for any experiment/proposal aiming to provide new results it is mandatory to reach a sensitivity of that level.
There is presently only one approved experiment for the Short--Baseline configuration with an accelerator beam, even if
previous proposals were available and new ones are under scrutiny, e.g. at JPARC~\cite{jparc}.

The sterile neutrino story has so far been developed either by trying to establish the hints (each at 2--3 $\sigma$ level) on
\nue appearance/disappearance, or looking at flavour connected channels, like the \numu disappearance one.
Within the next 2--3 years experiments on reactors and with radioactive sources can confirm or disprove the
\nue anomalies, while there is presently no reliable experiment~\cite{nessie} looking at the interference at the level of 1\% 
mixing between sterile and $\mu / \tau$
neutrino states other than the LBL ones. These  however have no possibility to observe the oscillation pattern.
Therefore, new specific  experiments should be settled and approved, in case reactor/source current expereriments would
provide positive results. 

\newpage



\Acknowledgements
I am grateful to Andrea Longhin and Stefano Dusini for providing critical suggestions to the manuscript.

\end{document}